\documentclass[aps,prl,twocolumn,floatfix,superscriptaddress]{revtex4}

\usepackage{graphicx}
\usepackage{lscape}
\usepackage{indentfirst}
\usepackage{latexsym}
\usepackage{multirow}
\usepackage{tabls}
\usepackage{epsfig}
\usepackage{multirow}

\usepackage{color}
\usepackage{amssymb}

\usepackage{amsfonts}
\usepackage{amsmath}
\usepackage{bm}
\usepackage{mathrsfs}

\def\vecmu{{\vec{\mu}}}

\def\lsim{\mathrel{\rlap{\lower3pt\hbox{\hskip1pt$\sim$}}
    \raise1pt\hbox{$<$}}}                
\def\gsim{\mathrel{\rlap{\lower3pt\hbox{\hskip1pt$\sim$}}
    \raise1pt\hbox{$>$}}}         

\def\coordeq{ \, \mathrel{ \rlap{\hbox{\hskip-2.5pt$=$} }
    \raise4pt\hbox{$\cdot$}} \, }                

\begin{document}

\title{Reduced basis catalogs for gravitational wave templates}

\def\addBrownPhys{Department of Physics, Brown University, Providence, RI 02912, USA}
\def\addCaltech{Jet Propulsion Laboratory and TAPIR, California Institute of Technology, Pasadena, CA 91125, USA}
\def\addUMD{Department of Physics, Maryland Center for Fundamental Physics, Joint Space Sciences Institute, Center for Scientific Computation and Mathematical Modeling, University of Maryland, College Park, MD 20742, USA}
\def\addBrownMath{Division of Applied Mathematics, Brown University, Providence, RI 02912, USA}
\def\addUWM{Center for Gravitation and Cosmology, University of Wisconsin-Milwaukee, Milwaukee, WI 53201, USA}
\def\addUMDEO{Department of Physics, Maryland Center for Fundamental Physics, University of Maryland, College Park, MD 20742, USA}

\author{Scott E. Field}
\affiliation{\addBrownPhys}

\author{Chad R.\,Galley}
\affiliation{\addCaltech} 
\affiliation{\addUMD}

\author{Frank Herrmann}
\affiliation{\addUMD}

\author{Jan S. Hesthaven}
\affiliation{\addBrownMath}

\author{Evan Ochsner}
\affiliation{\addUWM}
\affiliation{\addUMDEO}

\author{Manuel Tiglio}
\affiliation{\addUMD}

\begin{abstract}
We introduce a reduced basis approach as a new paradigm for modeling, representing and searching for gravitational waves.
We construct waveform catalogs for nonspinning compact binary coalescences, and we find that for accuracies of $99\%$ and $99.999\%$  
the method
generates a factor of about $10-10^5$ fewer templates 
than standard placement methods. The continuum of gravitational waves can be represented 
by a finite and comparatively compact basis. The method is robust under variations in the noise of detectors, implying that  
only a single catalog needs to be generated.
 
\end{abstract}

\maketitle 

{\it Introduction. }\, The second generation of earth-based gravitational wave detectors, such as Advanced LIGO and Virgo, will become operational in 2014-2015. These detectors are expected to directly measure gravitational waves (GWs), with likely event rates of $0.4-400$ per year for binary neutron stars (NS) and $0.4-1,000$  for binary black holes (BH)~\cite{Abadie:2010cfa}.
Direct detections would allow tests of general relativity in the nonlinear regime as well as access to portions of the universe otherwise unobservable.

Compact binary coalescences (CBCs), which consist of a pair of NSs and/or BHs inspiraling and merging, are considered to be one of the most promising sources of gravitational waves. The preferred method to search for GWs from CBCs is to employ matched filtering, which compares data from a detector to a bank of possible template waveforms and checks for a strong correlation between them.
In low mass searches, the inspiral regime dominates the observable signal but as the mass increases the merger regime becomes increasingly relevant. The merger regime requires numerical simulations -- even if used only for calibration of semianalytic models.
Given the number and cost of these simulations, knowing an optimal choice of parameters is critical  in order to limit the number of large simulations needed to accurately represent the variation over the parameter space.  
For this reason, it is desirable to seek a method that builds a template bank of waveforms by sequentially selecting only the most relevant points in the parameter space.

In addition, once a waveform catalog is constructed,  there is a significant computational cost in performing an actual search for GWs due to the size of the catalogs. 
Real-time analysis of the data is critical to generate alerts to search for electromagnetic counterparts and enable multimessenger astronomy \cite{Kanner:2008zh, Buskulic:2010zz,2010AAS...21540606S}. 
With the standard catalog placement method the number of templates needed grows rapidly with the dimension $P$ of the parameter space (as $(1-MM)^{-P/2}$, with $MM$ the minimal match \cite{Owen_B:96}) and such an approach could become impractical for searches of spinning binaries or other complex physical systems.

{\it Reduced Basis Method.}\, The RB framework~\cite{prud'homme:70} constructs a global basis rather than using local methods and can be seen as an application-specific spectral expansion. In such an approach one seeks to enable a rapid online evaluation of the reduced model at the expense of having to build the basis prior to the application. While such an approach is not suitable for certain types of applications, it is particularly well suited for those requiring near real-time or many inquiry response as is the case in the present application.  It has the following advantageous features over standard  model reduction techniques such as Proper Orthogonal Decompositions, Singular Value Decompositions, or Principal Component Analysis (see \cite{Pinnau2008} for a general review of these methods and \cite{Heng:2009zz,Cannon:2010qh} for applications to GWs), see also \cite{Brady:2004cf}:
\begin{enumerate}
\setlength{\itemsep}{1pt}\setlength{\parskip}{0pt}\setlength{\parsep}{0pt}
\item It is applicable to situations in which one must choose the most relevant parameters on the fly.
\item It yields nested, hierarchically constructed catalogs which can be easily extended. If $C_N = \{ h_1, \ldots , h_N\}$ is a catalog from the RB method then adding additional waveforms for higher accuracy implies that the resulting catalogs contain the previous ones, $C_N \subset C_{N+1} \subset C_{N+2} \cdots$.
\item It is highly computationally efficient. The cost of adding a new member to an existing catalog of size $N$ is independent of $N$. Hence, the total cost of generating a catalog of size $N$ scales linearly with $N$, in contrast to many other approaches.  
\item It yields catalogs that are nearly optimal in terms of the error in approximating the whole spectrum of GWs by a compact set of basis elements. This error is measured in the $L^{\infty}$ norm, ensuring a strict upper bound over the entire parameter space. 
\end{enumerate}

A gravitational wave is a function of time (or frequency in Fourier space) and of the $P$ parameters $\vecmu = \{ \mu_1, \ldots, \mu_P \}$ associated with the source. We denote each of them simply by $h_\vecmu$ and do not explicitly write the time or frequency dependence.
Let ${\cal H}$ be the space of all normalized GWs for the considered source(s). Although ${\cal H}$ is a not a linear space (the sum of two waveforms is not a waveform), we show that it can be represented by a linear one with arbitrarily high accuracy. We start with a theoretical description of our approach,  followed by a description of an actual implementation.

We are interested in approximating ${\cal H}$ by the best linear combinations of members $\Psi_i \equiv h_{\vecmu=\vecmu_i}$ of a catalog $C_N = \{ \Psi_i \}_{i=1}^N$. All such linear combinations form the {\em reduced basis space} $W_N = \mathrm{span}\left(C_N\right)$. The waveforms that make up this catalog could be optimally chosen so that the error in representing ${\cal H}$ with $W_N$ is minimized over the choice of $N$ catalog members. Such an optimal error is given by the Kolmogorov $N$-width~\cite{Pinkus},
\begin{align}
	d_N ({\cal H}) = \min_{C_N} \max_\vecmu \min_{u \in W_N} || u - h_\vecmu ||
	\label{Nwidth}
\end{align}
That is, one computes the error in the best approximation of $h_{\vecmu}$ by a member of $W_N$, then finds the parameter $\vecmu$ yielding the largest error, and lastly finds the smallest such error for all possible $N$-member catalogs. Here, the norm in Eq.\,(\ref{Nwidth}) is calculated from the complex inner product $\langle \cdot, \cdot \rangle$, which is related to the standard overlap of Wiener filtering by $4 \Re[\langle \cdot, \cdot \rangle]$, such that for two waveforms $F$ and $G$ in Fourier space,
\begin{align}
	\langle F, G \rangle \equiv \int_{f_L} ^{f_U}  \frac{ F^*(f) G(f) }{ S_n(f) } \, df
\end{align}
where $S_n(f)$ is the power spectral density (PSD) of the detector.

Finding a catalog that exactly achieves the $N$-width is a computationally demanding optimization problem. 
 Instead, we use a greedy approach, which is an inexpensive and practical procedure for hierarchically generating catalogs that \emph{nearly} satisfy the $N$-width~\cite{Binev10convergencerates}.

One constructs a catalog by first choosing a waveform for an arbitrary parameter value. A basis vector $e_1$ is then identified with this waveform, $e_1 = h_{\vecmu_1}$, and the catalog is $C_1 = \{ \Psi_1 = h_{\vecmu_1} \}$. To add another waveform to the catalog, one seeks the parameter value $\vecmu_2$ that maximizes $|| h_\vecmu - P_1 ( h_\vecmu) ||$ where $P_1 (h_\vecmu) =  e_1 \langle e_1, h_\vecmu \rangle$ is the (orthogonal) projection of $h_\vecmu$ onto $W_1$. We call this step a {\it greedy sweep}. The waveform corresponding to $\vecmu_2$ is added to the catalog so that $C_2 = \{ \Psi_1, \Psi_2 \}$. The new basis vector $e_2$ is then constructed via Gram-Schmidt orthonormalization. Notice that $C_1 \subset C_2$, which demonstrates the hierarchical nature of the catalogs generated. Additional members of the reduced basis catalog are generated by mathematical induction. At each step one picks up the parameter value $\vecmu_j$ that maximizes $|| h(\vecmu) - P_{j-1}(h(\vecmu)) ||$.

\begin{figure}
\includegraphics[width=\columnwidth]{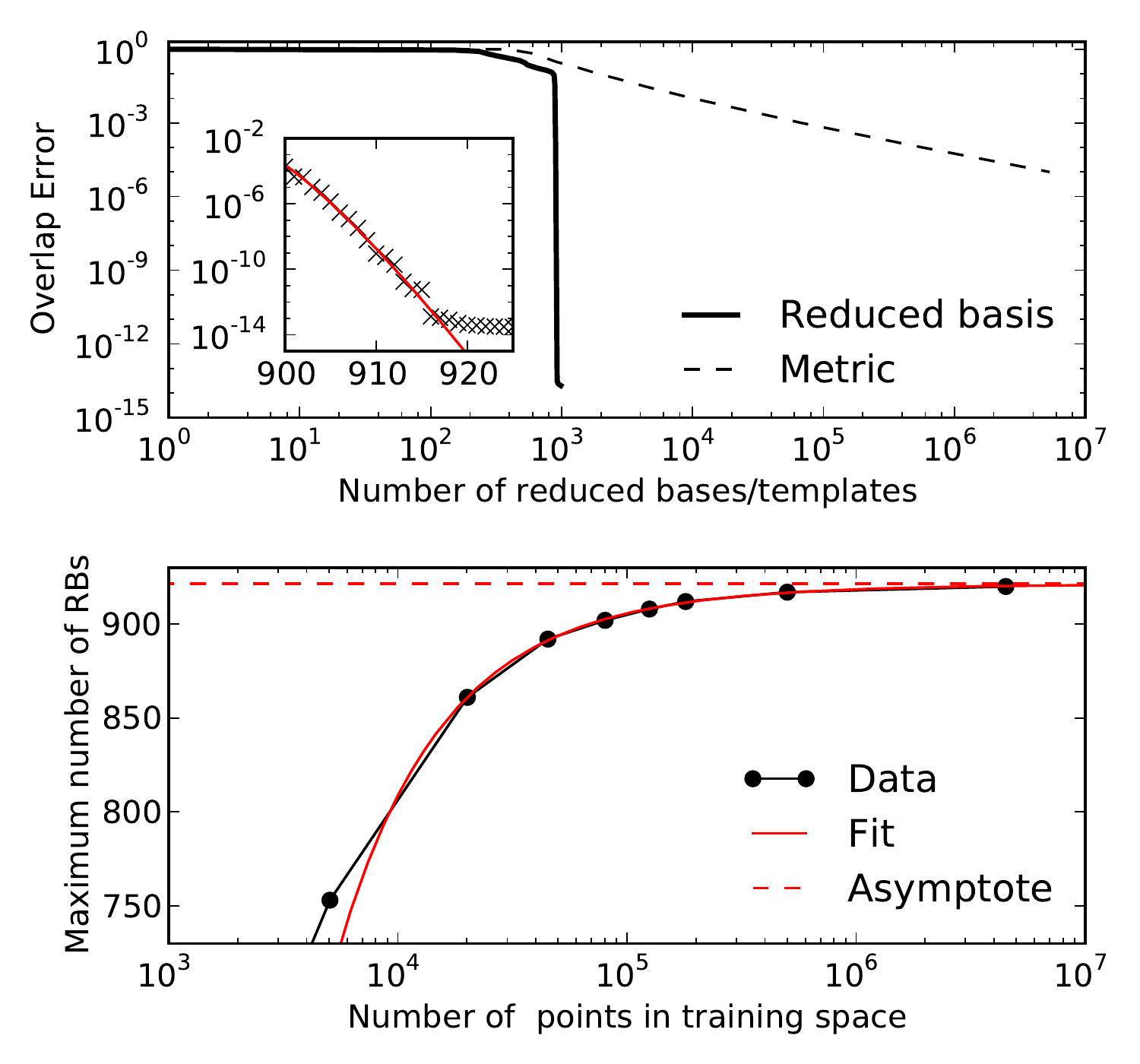}
\caption{Error in approximating the space of waveforms by a discrete catalog for BNS inspirals with Initial LIGO.  For reduced basis, the error is the square of the  greedy error (\ref{eq:error_greedy}) while for metric placement the error is $(1-MM)$ with $MM$ the minimal match. The lower panel shows the extrapolation of the maximum number of RBs generated for an infinitely large training space. The fit shown (red) excludes the two points with largest $x$, which change the asymptotic value by $0.2$.}
\label{fig:errors}
\end{figure}

It can be shown \cite{Binev10convergencerates} that if the decay of the $N$-width with $N$ can be bounded by an exponential, 
$$
d_N({\cal H}) \leq A e^{- c  N^{\alpha}} ,
$$
for some real $c$ and $\alpha$, then the decay of the maximum error for a catalog $C_N$ generated by this approach, which we call the \emph{greedy error} $\varepsilon_N$, is also exponential, 
\begin{equation} 
	\varepsilon_{N} \equiv \max_\vecmu || h_\vecmu - P_N(h_\vecmu) || \leq \tilde{A} e^{- d N^{\beta}} 
	\label{eq:error_greedy}
\end{equation}
where $P_N(h_\vecmu)=\sum_{i=1}^{N}\langle e_i,h_\vecmu \rangle e_i$ and $d, \beta$ depend on $c,\alpha$ (see \cite{Binev10convergencerates} for more details).
Similar results hold in the case of power-law fall-off, i.e. $d_N({\cal H}) \leq BN^{-s}$ implies $\varepsilon_{N}\leq \tilde{B}N^{-s}$.
Note that $\varepsilon_{N}$ is a bound on the error between a waveform and its representation, and that $\varepsilon_{N}^2 = \max_\vecmu \left( 1-\Re[\langle h_\vecmu,P_N(h_\vecmu)\rangle] \right)$, so that $\varepsilon_{N}^2$ is an error on the overlap directly comparable to $(1-MM)$.
Given that GWs appear to depend smoothly on the parameters $\vecmu$, we expect $d_N({\cal H})$, and hence the greedy error $\varepsilon_{N}$, to decay rapidly (in fact exponentially) with $N$, which is a key feature of this method. Notice that (\ref{eq:error_greedy}) implies that any waveform can be represented as $h_\vecmu = P_N(h_\vecmu) + \delta h_\vecmu (f)$ where $|| \delta h_\vecmu (f) || \le \varepsilon_N$. Therefore, if $\varepsilon_N$ is of the order of numerical round-off then, in practice, the projection of $h_\vecmu$ onto $W_N$ equals the waveform itself. In addition, the number of RBs needed to represent any $h_\vecmu$ is comparatively small (see below).

In the case in which one is numerically solving equations on the fly while building the RB, the error (\ref{eq:error_greedy}) is replaced by an inexpensive error estimate evaluation (such as a residual), which is referred to as the {\it weak greedy approach}. Once the next parameter value is chosen one solves the full problem for it (referred to as the {\it offline stage}) and proceeds to the next greedy sweep. In any greedy approach, the maximum over $\vecmu$ is searched for, in practice, using a {\it training space} of samples $\vecmu$.  However, since this is done as part of the offline process, the training space can be finely sampled and one can take advantage of the observation that  evaluations for different parameters values are  decoupled and, hence, embarrassingly parallel. 

If one attempted a matched filter search with a RB catalog $C_N$ by filtering each basis function against the data and maximizing over arbitrary linear combinations of these filter outputs, one would of course get a very high false alarm rate. Instead, it is important to allow only linear combinations that correspond to physical waveforms. To be more specific, one could easily store the matrix of overlaps between
waveforms in the original template bank (the training space) and the reduced basis, i.e.,
$\alpha_{ij} = \langle e_i , h_{\vecmu_j} \rangle$. In fact, our algorithm provides this
reconstruction matrix as output. A matched filtering computation may then be performed by integrating the incoming signal $s$ against each member
of the basis $\langle s , e_i \rangle$ and using the reconstruction matrix $\alpha_{ij}$ to
recover the matched filtering integral of the signal with any template $h_{\vecmu_j}$ in
the original bank. Explicitly, $\langle s , h_{\vecmu_j} \rangle = \sum_i \langle s, e_i \rangle \alpha_{ij}$.
In this way, using the reduced bases is equivalent to using the original waveform space but with
many fewer matched filtering integrals to compute for a given signal. Hence, using RB yields no increase in the false alarm rate.

{\it Catalogs for compact binary inspirals.}\, We discuss our results for constructing reduced bases for ``chirp'' gravitational waveforms for binary inspirals without spins \cite{Blanchet:1995fg,Will:1996zj}. We use the 2nd order post-Newtonian accurate waveforms in the stationary phase approximation, which are known in closed form, so that the parameter space is two-dimensional (the binary's masses).
For simplicity, we take the coalescence time and phase to be constant for each waveform.

Fig.~\ref{fig:errors} shows results for the greedy error using a reduced basis model  for inspirals of binary neutron stars (BNS) with mass components in the range [1-3]$M_{\odot}$ (for Initial LIGO with a lower frequency cutoff at $40$ Hz) compared with the standard metric template placement method~\cite{Owen_B:96}. After a slowly decaying region, the reduced basis model gives very fast exponential convergence decay, which can be fitted by $\varepsilon_{N}^2 = ae^{-b N^p}$ with $a=9.65 \times 10^{-4}$, $b=0.598$, $p=1.25$. The metric method yields approximately linear decay for a two-dimensional parameter space. As already mentioned, this decay becomes slower as the dimensionality $P$ of the parameter space increases. The fast decay of the reduced basis model allows a representation of the whole set of gravitational waves for these sources and mass ranges to within machine precision. We have found the same feature in all mass ranges that we have explored. This leads to the rather remarkable finding that for all practical purposes the set of relevant gravitational waveforms in compact parameter regions appears to be {\em finite} dimensional. When increasing the number of samples $x$ in the training set we find the following fit for the number of RB for machine precision error, $N= a+b x^{-1/2} + c x^{-1}$ with $a=921, b=-2090, c=-9.18 \times 10^5$ for the case of Fig.~\ref{fig:errors}. In particular, in the limit $x \rightarrow \infty$ only $921$ bases are needed to represent, within numerical accuracy, the full space of waveforms ${\cal H}$ for this range of masses for BNS inspirals. 

\begin{figure}
\includegraphics[width=\columnwidth]{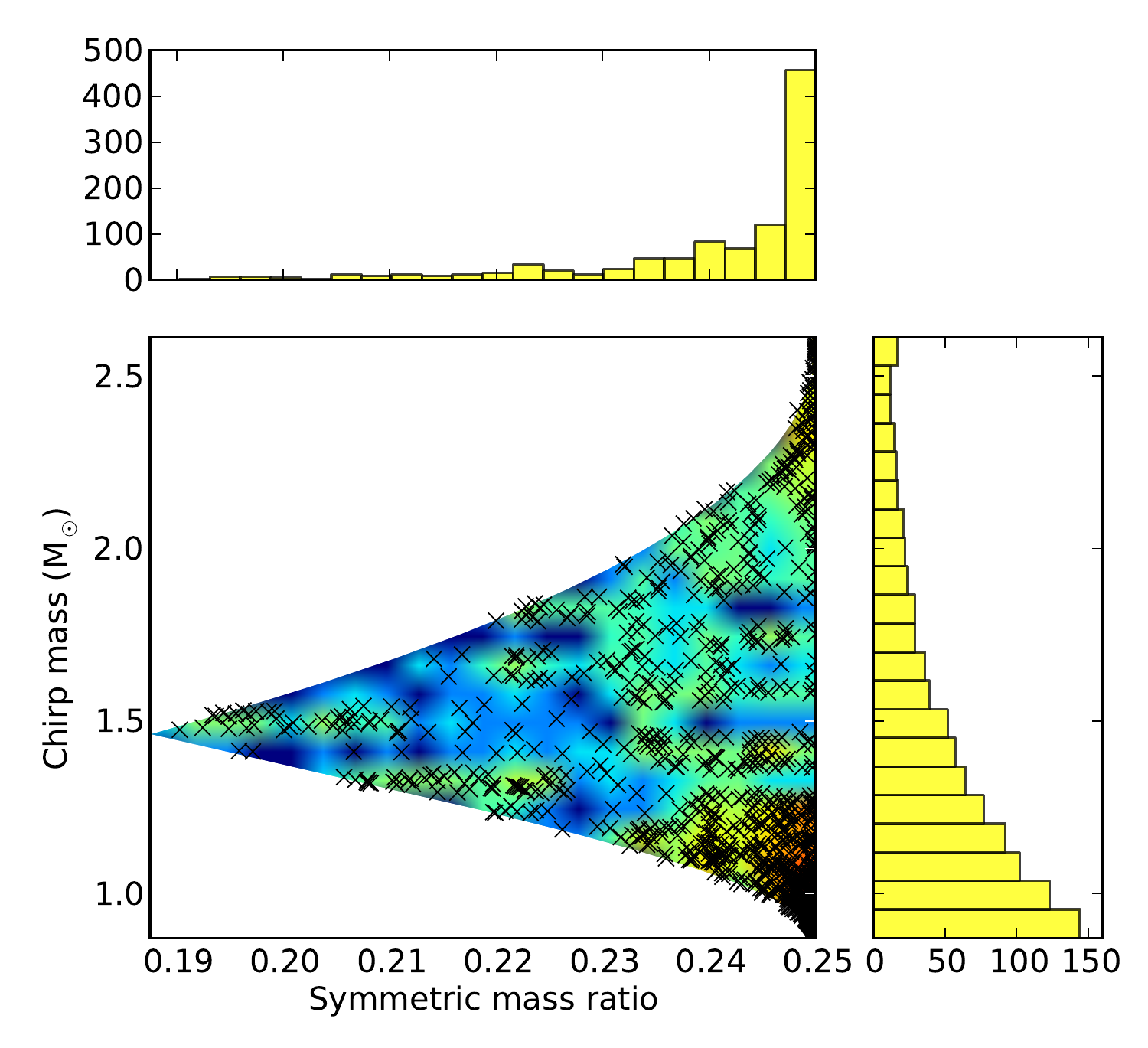}
\caption{The points show the parameter values chosen for the catalog of BNS and Initial LIGO. The density of parameter values is shown using a coloramp as well as histograms.}
\label{fig:params}
\end{figure}

Figure~\ref{fig:params} shows the chosen parameter values in the chirp mass vs symmetric mass ratio plane and a density plot of the number of RBs. The histograms highlight that most values are picked for (nearly) equal mass systems of low chirp mass.

Table~\ref{tab:Nrb} shows the number of RB that we need to represent, for different overlap error tolerances, inspirals of BNS and stellar size binary black holes (BBH, with mass components in  the range [3-30]$M_{\odot}$). 
The limit $x \rightarrow \infty$ is not taken here for simplicity so the RB values listed in Table~\ref{tab:Nrb} are slightly underestimated.  
 
\begin{table}[t]
	\caption{Number of reduced bases/templates for different target accuracies with the reduced basis (RB) and template metric (TM) approaches for binary neutron stars (BNS) and binary black holes (BBH), using spinless chirp waveforms. We assume a lower frequency cutoff of $40$ Hz for Initial LIGO and $10$ Hz for Advanced LIGO and Virgo. The error is given by $\varepsilon_{N}^2$ for RB and $(1-MM)$ for TM. }
\begin{tabular}{|c|c|c|c|c|c|}
	\hline
	\multirow{2}{*}{Detector} & \multirow{1}{*}{Overlap} & \multicolumn{2}{|c}{BBH} & \multicolumn{2}{|c|}{BNS} \\ 
	 	\cline{3-6}
	\multirow{2}{*}{~} & \multirow{1}{*}{Error} & RB & TM & RB & TM \\
	\hline\hline
	\multirow{3}{*}{InitLIGO} & $10^{-2}$ & $165$ & $2,450$ & $898$ & $10,028$ \\
		& $10^{-5}$ & $170$ & $ 1.2 \times 10^6    $ & $904$ & $4.3 \times 10^6$ \\
		& $2.5 \times 10^{-13}$ & $182$ & $ 5.9 \times 10^{12} $ & $917$ & $ 1.4 \times 10^{13} $ \\
	\hline
	\multirow{3}{*}{AdvLIGO} & $10^{-2}$ & $1,058$ & $19,336$ & $5,395$ & $72,790$ \\
		& $10^{-5}$ & $1,687$ & $ 1.5 \times 10^7   $ & $8,958$ & $4.9 \times 10^7 $ \\
		& $2.5 \times 10^{-13}$ & $1,700$ & $  2.3 \times 10^{14} $ & $8,976$ & $ 5.6 \times 10^{14}$ \\
	\hline
	\multirow{3}{*}{AdvVirgo} & $10^{-2}$ & $1,395$ & $42,496$ & $7,482$ & $156,127$ \\
		& $10^{-5}$ & $1,690$ & $3.1 \times 10^7$ & $8,960$ & $8.3 \times 10^7 $ \\
		& $2.5 \times 10^{-13}$ & $1,703$ & $ 4.8 \times 10^{14} $ & $8,977$ & $6.0 \times 10^{14}$  \\
	\hline
\end{tabular}
\label{tab:Nrb}
\end{table}

{\it Sensitivity to Nonstationary  Noise.}\, The PSD of any ground-based interferometer will fluctuate in time due to changes in environmental noises and other factors. Since the PSD weights the inner products used to construct the reduced basis, one might worry that a new RB needs to be constructed for any variation in the PSD. 

Remarkably, we find indications that the RB constructed assuming a fiducial PSD is highy robust against rather large perturbations. From a histogram of the sensitivity of the LIGO interferometers during a portion of LIGO's fifth science reported in~\cite{Abbott:2007kv}, we conclude that a $20\%$ increase or decrease in the sensitivity to BNS signals is a large but realistic fluctuation to the sensitivity of the Initial LIGO interferometers. Therefore, we constructed smooth deformations of the Initial LIGO design PSD meant to simulate variations in the seismic, thermal and shot noise levels which yield roughly a $20\%$ increase or decrease in sensitivity to BNS signals. We find that the RB generated for the Initial LIGO design PSD with a greedy error tolerance of $\varepsilon_N$ can represent any inspiral waveform within the perturbed PSDs considered here with an error of no more than $1.3 \varepsilon_N$.
This result implies that one needs to compute only a {\em single reduced basis} for a given source for a particular detector. This is unlike the current operating procedure in which a new template bank is generated every $\sim 2048$ seconds because of the drifts in the nonstationary noise. We will provide further details in a forthcoming paper.

{\it Conclusions.}\, We have considered the development and use of a reduced basis method to template bank construction and found rapid exponential convergence of the waveform catalog over the full parameter space. The catalog is computationally cheap to derive, hierarchical (i.e. if a more accurate catalog is required, elements can be added), can be extended for a computational cost that is independent of $N$, and is robust under changes in a detector's noise. We have found that the space of gravitational waveforms considered in this paper is essentially finite-dimensional for any finite range of physical parameters, and conjecture that it is in general the case. 
Elsewhere, we will present a more detailed description of these results and further applications of the RB framework.

{\it Acknowledgments.}\, We thank P. Ajith, E. Barausse, A. Buonanno, C. Cutler, A. Le Tiec, S. Lau, T. Littenberg, J. McFarland, R. Nochetto, L. Price, M. Vallisneri and B. Zheng for helpful discussions and suggestions, and especially B. Stamm for introducing several of us to the use of Reduced Basis. This work has been supported by NSF Grants PHY0801213 and PHY0908457 to UMD, NSF DMS 0554377 and OSD/AFOSR FA9550-09-1-0613 to Brown Univ. and NSF Grant PHY0970074 to  UWM. C.\,G. was supported in part by an appointment to the NPP at JPL through a contract with NASA. 

\bibliographystyle{physrev}
\bibliography{references}

\end{document}